\documentclass[aps,prd,twocolumn,superscriptaddress,nofootinbib,preprintnumbers]{revtex4-1}
\usepackage{amsmath}
\usepackage{graphicx}
\usepackage{subfigure}
\usepackage{amssymb}
\usepackage{xcolor}
\usepackage{multirow}
\usepackage{cancel}
\usepackage{color}
\usepackage{ulem}
\usepackage{listings}
\usepackage{xcolor}
\usepackage{float}
\usepackage{slashed}

\usepackage[colorlinks,citecolor=blue]{hyperref}
\usepackage{amsmath}%\ge=\geqslant; \le=\leqslant
\usepackage{wrapfig}

\newcommand{\be}{\begin{equation}}
\newcommand{\ee}{\end{equation}}
\newcommand{\beq}{\begin{equation}}
\newcommand{\eeq}{\end{equation}}
\newcommand{\bea}{\begin{eqnarray}}
\newcommand{\eea}{\end{eqnarray}}
\newcommand{\besp}{\begin{equation}\begin{split}}
\newcommand{\eesp}{\end{split}\end{equation}}

\newcommand{\nn}{\nonumber}

\newcommand{\Eq}[1]{Eq.~(\ref{#1})}
\newcommand{\Dfbd}{\mathord{\buildrel{\lower3pt\hbox{$\scriptscriptstyle\leftrightarrow$}}\over {D}_{\mu}}}
\newcommand{\ave}[1]{\left\langle #1\right\rangle}

\def\mL{\mathcal{L}}

\def\mO{\mathcal{O}}

\def\Z{\mathbb{Z}}

\def\0{\textbf{0}}
\def\1{\textbf{1}}
\def\2{\textbf{2}}
\def\3{\textbf{3}}
\def\4{\textbf{4}}
\def\5{\textbf{5}}
\def\6{\textbf{6}}
\def\7{\textbf{7}}
\def\8{\textbf{8}}
\def\9{\textbf{9}}

\def\p{\textbf{p}}

\begin{document}
\rightline{CTPU-PTC-20-20}

\title{Fermi-ball dark matter from a first-order phase transition}

\author{Jeong-Pyong Hong}
\email{jeongpyonghong@ibs.re.kr}
\affiliation{Center for Theoretical Physics of the Universe, Institute for Basic Science (IBS), Daejeon 34126, Korea}
\affiliation{Center for Theoretical Physics, Department of Physics and Astronomy, Seoul National University, Seoul 08826, Korea}

\author{Sunghoon Jung}
\email{sunghoonj@snu.ac.kr}
\affiliation{Center for Theoretical Physics, Department of Physics and Astronomy, Seoul National University, Seoul 08826, Korea}

\author{Ke-Pan Xie}
\email{\text{Corresponding author: }kpxie@snu.ac.kr}
\affiliation{Center for Theoretical Physics, Department of Physics and Astronomy, Seoul National University, Seoul 08826, Korea}

\begin{abstract}

We propose a novel dark matter (DM) scenario based on a first-order phase transition in the early universe. If dark fermions acquire a huge mass gap between true and false vacua, they can barely penetrate into the new phase. Instead, they get trapped in the old phase and accumulate to form macroscopic objects, dubbed Fermi-balls. We show that Fermi-balls can explain the DM abundance in a wide range of models and parameter space, depending most crucially on the dark-fermion asymmetry and the phase transition energy scale (possible up to the Planck scale). They are stable by the balance between fermion's quantum pressure against free energy release, hence turn out to be macroscopic in mass and size. However, this scenario generally produces no detectable signals (which may explain the null results of DM searches), except for detectable gravitational waves (GWs) for electroweak scale phase transitions; although the detection of such stochastic GWs does not necessarily imply a Fermi-ball DM scenario.
 
\end{abstract}

\maketitle

\section{Introduction}

The particle origin of dark matter (DM) is a long-standing mystery. Cosmological observations show that DM contributes $\sim$27\% of the total energy of the universe~\cite{Aghanim:2018eyx}, but none of the Standard Model (SM) particles can serve as DM candidates~\cite{Bertone:2004pz}. New weakly interacting massive particles (WIMPs)~\cite{Lee:1977ua} with the freeze-out mechanism has been the most popular explanation for DM for several decades. However, the continuously reported null results from the direct~\cite{Schumann:2019eaa}, indirect~\cite{Gaskins:2016cha} and collider~\cite{Boveia:2018yeb} searches motivate new DM paradigms beyond WIMPs.

Recently, there has been a growing number of studies on the DM generated in association with a first-order cosmic phase transition (FOPT). During a FOPT, the discontinuity of the scalar vacuum expectation value (VEV) could be crucial in DM physics, by altering the decay of DM~\cite{Baker:2016xzo,Baker:2018vos,DiBari:2020bvn}, by generating asymmetric DM~\cite{Kaplan:1991ah,Dutta:2006pt,Dutta:2010va,Shelton:2010ta,Petraki:2011mv,Walker:2012ka,Baldes:2017rcu,Gu:2017rzz,Hall:2019rld}, by producing DM non-thermally~\cite{Falkowski:2012fb}, by filtering DM to the true vacuum~\cite{Baker:2019ndr,Chway:2019kft,Marfatia:2020bcs}, by condensing particles into the false vacuum to form (scalar) $Q$-ball DMs~\cite{Krylov:2013qe,Huang:2017kzu} or quark (or quark-like fermion) nuggets~\cite{Witten:1984rs,Bai:2018vik,Bai:2018dxf,Zhitnitsky:2002qa,Atreya:2014sca,Oaknin:2003uv,Lawson:2012zu,Frieman:1990nh}, and by producing the primordial black holes~\cite{Dymnikova:2000dy,Khlopov:2008qy}~\footnote{FOPTs can also happen in freeze-out~\cite{Alanne:2014bra,Fairbairn:2013uta,Li:2014wia,Petraki:2007gq,Chung:2011hv,Chung:2011it,Chao:2017vrq,Jiang:2015cwa,Liu:2017gfg,Jaramillo:2020dde} or freeze-in~\cite{Cohen:2008nb,Baker:2017zwx,Bian:2018mkl,Bian:2018bxr} processes, potentially changing the thermal history of DM.}.

\begin{figure}
\centering
\subfigure{
\includegraphics[scale=0.32]{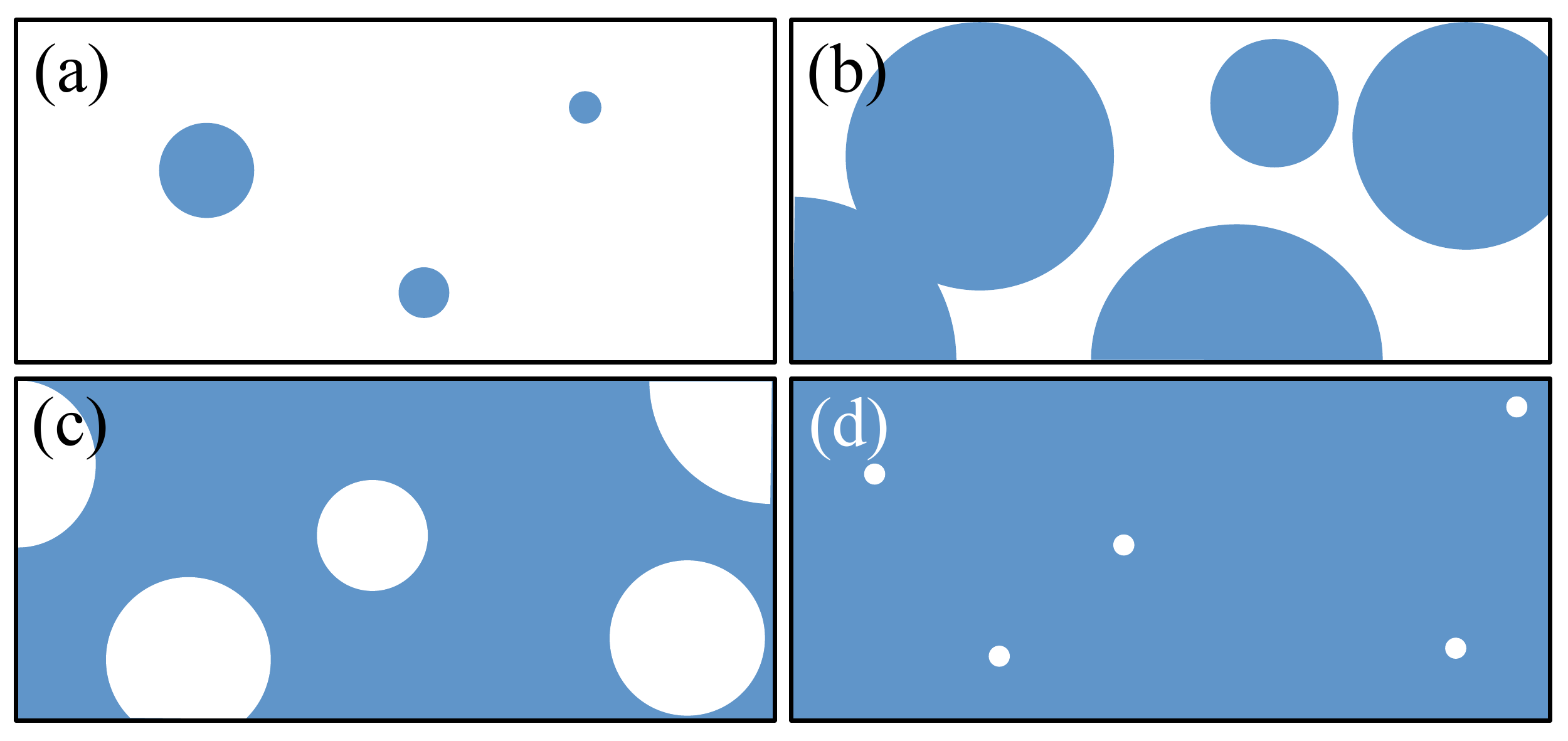}}
\caption{Sketch of the mechanism. {\bf (a), (b)} True vacuum bubbles (blue) emerge in the false vacuum background (white) and grow up, collide and fulfill more and more space. {\bf (c)} The true vacuum has dominated the universe, however there is still an $\mO(1)$ fraction of volume occupied by the false vacuum; fermions and anti-fermions are trapped in those remnants. {\bf (d)} Vacuum pressure shrinks the false vacuum remnants into tiny objects, and all anti-fermions are annihilated away. Finally, stable Fermi-balls with net $Q$-charge are formed.}
\label{fig:sketch}
\end{figure}

In this article, we propose a new mechanism in which during a FOPT dark {\it fermions} are trapped inside the false vacuum to subsequently form compact macroscopic DM candidates, which we call ``Fermi-balls''. This scenario requires the following three conditions to be satisfied:
\begin{enumerate}
\item First, the fermion field needs to have a huge mass gap between the false and true vacua compared to the phase transition temperature, so that it cannot penetrate into the true vacuum due to energy conservation, hence are trapped in the false one. 
\item Second, there should be an asymmetry of the number density between dark fermions and anti-fermions, so that excess fermions survive from pair annihilations, congregate and construct Fermi-balls. 
\item Third, the fermion field should carry a conserved global $U(1)_Q$ charge so that the Fermi-ball accumulates a net $Q$-charge ensuring its stability. 
\end{enumerate}
Each condition is met in a wide varieties of new physics models. The mechanism is illustrated in Fig.~\ref{fig:sketch}~\footnote{We use ``bubbles'' to represent objects that contain the true vacuum, while ``remnants'' for objects with the false vacuum.}.

The Fermi-ball has several novelties compared to other similar mechanisms. First, it is made of fermions, while a $Q$-ball~\cite{Kusenko:1997si} -- which also localizes conserved charges in small objects -- is made of scalars. This leads to different profiles as will be shown. Although fermions can also be trapped by domain walls~\cite{Lee:1986tr,Holdom:1987ep,Macpherson:1994wf} so that the surface tension dominates the free energy release suppressed by fine-tuned degenerate minima, the Fermi-ball formation is dominated by large free energy release which is more natural in strong FOPTs. 

In Section~\ref{sec:setup}, we present a basic model framework realizing all three conditions. Based on it, intermediate physics of the Fermi-ball formation is discussed, starting from the trapping in Section~\ref{sec:trapping}, then supercooling in \ref{sec:supercooling}, stable Fermi-ball formation in \ref{sec:profile}, and finally Fermi-ball DM properties in \ref{sec:FBDM}. We demonstrate the Fermi-ball scenario using a toy model with example full numerical results in Section~\ref{sec:toy}, and investigate possible detectable signals of the gravitational waves (GWs) in Section \ref{sec:GW}. We summarize in Section~\ref{sec:conclusion}. $\hbar = c =1$ is adopted throughout this paper.

%%%%%%%
\section{Basic setup} \label{sec:setup}

Consider a real scalar field $\phi$. Its thermal potential $U(\phi,T)$  triggers a FOPT from $\ave{\phi}=0$ to $\ave{\phi}= w(T)$  at a temperature $T$ below the critical temperature $T_c$. During the FOPT, vacuum bubbles containing the new phase emerge and expand in the universe, and $\ave{\phi}$ varies smoothly from $w(T)$ to 0 on the bubble wall. A more detailed description of (supercooling) FOPTs relevant to this work will be provided in Section~\ref{sec:supercooling}. 

Let $\chi$ be a dark Dirac fermion with the Lagrangian
\be \label{L_chi}
\mL\supset\bar\chi i\slashed{\partial}\chi-g_\chi\phi\bar\chi\chi,
\ee
which enjoys a global $U(1)_Q$ symmetry with the $Q$-charge $+1$ for $\chi$. \Eq{L_chi} implies that $\chi$ is massless in the false vacuum, while acquiring a mass $M_\chi(T)=g_\chi w(T)$ in the true vacuum. The first condition in the Introduction is expressed as
\be\label{large_mass}
M_\chi^*= g_\chi w_*\gg T_*,
\ee
which makes most $\chi$'s fail to penetrate into the true vacuum because their average kinematic energy is $\mO(T_*)$. The subscript $*$ implies that the parameters and the condition are defined at $T_*$ at which Fermi-balls start to form; $T_*$ is below $T_c$ and will be defined later. \Eq{large_mass} can be realized by either large $w_*/T_*$ (supercooling)~\cite{Creminelli:2001th,Nardini:2007me,Konstandin:2011dr,Jinno:2016knw,Marzo:2018nov} or strong $g_\chi \gg 1$~\cite{Carena:2004ha,Angelescu:2018dkk}; but as will be shown, the supercooling with $g_\chi \sim {\cal O}(1)$ is good enough for Fermi-ball DM scenarios. Examples of dark fermions acquiring a large mass in the true vacuum can be found in Refs.~\cite{Hui:1998dc,Hambye:2018qjv,Baratella:2018pxi}.

The $U(1)_Q$ symmetry of \Eq{L_chi} ensures the stability of Fermi-balls with a net $Q$-charge (the third condition); in other words, $\chi$ does not decay. However, if this were the exact symmetry at all energy scales, there could not have been an asymmetry between $\chi$ and $\bar\chi$ so that no $\chi$ can survive from the pair annihilation $\chi\bar\chi\to\phi\phi$ when the false vacuum remnants shrink (the second condition). The key point of the solution is that the $U(1)_Q$ is a good symmetry at low energy/temperature while it was broken at some high energy scale. At this high energy, the asymmetry can be generated by various asymmetric DM mechanisms~\cite{Kaplan:2009ag,Petraki:2013wwa,Zurek:2013wia}, independently from other sectors of the model; one concrete example similar to the leptogenesis is presented in Appendix~\ref{sec:excess}. In the main part of this work, it is just good enough to parametrize the $\chi$-asymmetry as
\be
\eta_\chi=c_\chi \eta_B,
\ee
where $\eta_B\sim10^{-10}$ is the baryon asymmetry, and $c_\chi$ is a free parameter which can be easily very small, at least ${\cal O} (0.01)$ relevant to the toy model in this work; see Appendix~\ref{sec:excess}.

For Fermi-balls to be abundantly formed, $\chi$ must be in the thermal bath before the FOPT. Provided that $g_\chi$ is not feeble (always satisfied by the stability condition, as will be shown), $\chi$ can be in equilibrium with $\phi$ via $\chi\bar\chi\leftrightarrow\phi\phi$. Further, $\phi$ can be in equilibrium with the SM particles via scalar portal couplings, e.g. $\phi^2 h^2$~\cite{Silveira:1985rk,Burgess:2000yq,Patt:2006fw} with $h $ being the SM Higgs. In the meantime, those couplings also make $\phi$ disappear from the false vacuum after the FOPT, e.g. through pair annihilations $\phi \phi \to h h$ or $\phi \phi \to f \bar{f}$ to SM fermions or through decays $\phi\to hh$ or $\phi\to f\bar f$, so that only $\chi$'s survive and accumulate in the false vacuum. The $\phi$-portal coupling is assumed to be strong enough for these to happen, while weak enough not to affect the FOPT and SM-like Higgs couplings~\cite{Baker:2019ndr}.

With these basic setups and ingredients, we discuss Fermi-ball physics in the following sections, later with full numerical results for a toy model.

%%%%%%%
\section{Trapping fermions in false vacuum} \label{sec:trapping}

The trapping efficiency can be calculated by investigating the kinematics around the expanding bubble wall. The free energy difference $\Delta U_*$ between the true and false vacua pushes the bubble to expand, while the reflection of particles on the wall acts as a pressure $P$ that tends to stop the expansion. When the balance between them $P=\Delta U_*$ is achieved, the bubble reaches its terminal velocity $v_b$. It can be solved numerically for a given model~\cite{Chway:2019kft,Bodeker:2009qy,Bodeker:2017cim,Hoeche:2020rsg,Ellis:2019oqb}, but in this paper it will be treated as a free parameter of ${\cal O}(0.1)$. If the bubble radius is much larger than wall thickness, the vicinity of the wall can be treated as a one-dimensional problem: the wall is parallel to the $Oxy$ plane and moving along the $z$-axis with the velocity $v_b$. 

Outside the bubble is the false vacuum (f.v.), in which $\chi$ is massless and in thermal equilibrium, i.e.
\be\label{f_chi}
f_\chi^{\rm f.v.}(\p)=\frac{1}{e^{(|\p|-\mu_\chi)/T_*}+1},
\ee
where $\mu_\chi$ is the chemical potential. The number density of $\chi$ is given by
\be\label{n_chi}
n_\chi^{\rm f.v.}=2\int\frac{d^3\p}{(2\pi)^3}f_\chi^{\rm f.v.}(\p)\approx\frac{3\zeta(3)}{2\pi^2}T_*^3+\frac{\mu_\chi}{6}T_*^2.
\ee
In the wall rest frame, the $\chi$ distribution is
\be
\tilde f_\chi^{\rm f.v.}(\p)=\frac{1}{e^{(\gamma_b|\p|+\gamma_bv_bp_z-\mu_\chi)/T_*}+1},
\ee
where $\gamma_b=(1-v_b^2)^{-1/2}$ is the Lorentz factor. Only $\chi$ with $-p_z>M_\chi^*$ can pass across the wall, due to the energy conservation. The particle current per unit area and unit time is then~\cite{Chway:2019kft}
\be
\tilde J_\chi=2\int\frac{d^3\p}{(2\pi)^3}\frac{-p_z}{|\p|}\tilde f_\chi^{\rm f.v.}(\p)\Theta(-p_z-M_\chi^*),
\ee
where $\Theta$ is the Heaviside step function. The particle current can be transformed into the plasma frame by multiplying a time dilation factor $J_\chi=\tilde J_\chi/\gamma_b$. The number density of $\chi$ penetrating into the true vacuum is $n_\chi^{\rm pene.}=J_\chi/v_b$. This derivation is valid only when $n_\chi^{\rm pene.}\ll n_\chi^{\rm f.v.}$ such that the $\chi$ in the false vacuum can be approximated to be in equilibrium.

\begin{figure}
\centering
\subfigure{
\includegraphics[scale=0.35]{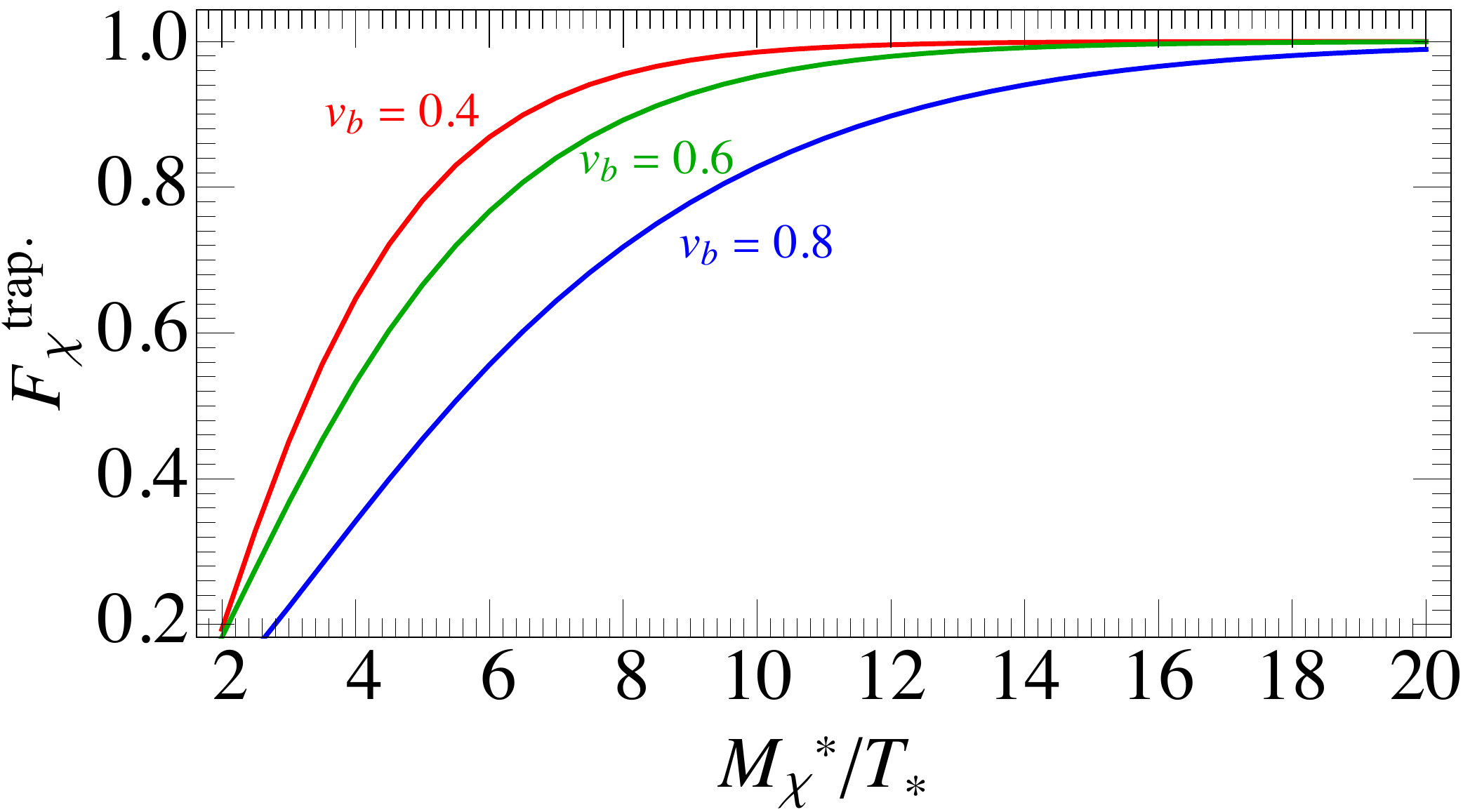}}
\caption{The fraction of $\chi$ trapped inside the false vacuum.}
\label{fig:trapping}
\end{figure}

The fraction of $\chi$ trapped in the false vacuum is
\be\label{trap}
F_\chi^{\rm trap.}=1-\frac{n_\chi^{\rm pene.}}{n_\chi^{\rm f.v.}}.
\ee
The trapping fraction of $\bar\chi$ can be obtained by replacing $\mu_\chi\to-\mu_\chi$. But the difference is negligible during trapping. If the $\chi$-asymmetry (from \Eq{n_chi})
\be
\eta_\chi\equiv \left(n_\chi^{\rm f.v.}-n_{\bar\chi}^{\rm f.v.}\right)/s_*=\frac{15\mu_\chi}{2\pi^2g_*T_*},
\ee
with $s_*=2\pi^2g_*T_*^3/45$ and $g_*$ being the entropy density and number of relativistic degrees of freedom at $T_*$ respectively, is not far away from $\eta_B = 0.9\times10^{-10}$~\cite{Tanabashi:2018oca}, $g_* \sim100$ yields small $\mu_\chi/T_*\sim10^{-8}$, thus can be ignored. Therefore, $F_\chi^{\rm trap.}\approx F_{\bar\chi}^{\rm trap.}$.

Fig.~\ref{fig:trapping} shows the trapping fraction as a function of $M_\chi^*/T_*$ for different $v_b$. For a reasonably large $M_\chi^*/T_* \gtrsim {\cal O}(1)$ and relativistic $v_b \sim {\cal O}(0.1)$, the trapping is very efficient with the fraction close to 100\%. For a given $M_\chi^*/T_*$, the fraction decreases with $v_b$ because $\chi$ in the wall frame becomes more energetic, having higher probability to penetrate the mass barrier.

%%%%%%%
\section{The Fermi-ball DM} \label{sec:DM}

%%%%%
\subsection{Formation of Fermi-ball} \label{sec:supercooling}

Starting from the general description of the steps of a FOPT, we define characteristic temperatures including the Fermi-ball formation temperature $T_*$. 
Generally speaking, a FOPT is the vacuum decay between two local minima, $\ave{\phi}=0$ and $\ave{\phi}=w(T)$. At the critical temperature $T_c$, the two minima are degenerate, separated by a barrier. As temperature falls below $T_c$, the $\ave{\phi}=w(T)$ phase becomes the new global minimum (true vacuum) and the universe starts decaying to it. The decay rate per unit volume reads~\cite{Linde:1981zj}
\be
\Gamma(T) \, \approx \, T^4e^{-S_3(T)/T},
\ee
with $S_3(T)$ being the action of the $O(3)$ symmetric bounce solution.

Once the vacuum transition probability in a Hubble volume and a Hubble time reaches $\mO(1)$, the bubble nucleation becomes efficient
\be\label{nucleation_condition}
\Gamma(T_n)H^{-4}(T_n) \, \approx\,1,
\ee
defining the nucleation temperature $T_n$. $H(T)$ is the Hubble constant 
\be\label{supercooling_H_sim}
H^2(T)=\frac{8\pi}{3M_{\rm Pl}^2}\left(\frac{\pi^2}{30}g(T) T^4+\Delta U(T)\right),
\ee
with $M_{\rm Pl}=1.22\times10^{19}~{\rm GeV}$ being the Planck scale, $g(T)$ being the number of relativistic degrees of freedom, and $\Delta U(T)$ being the (positive definite) potential energy difference between true and false vacua
\be
\Delta U(T)=U(0,T)-U\left(w(T),T\right).
\ee
The $\Delta U(T)$ in \Eq{supercooling_H_sim} is usually ignored because normally such energy release from a phase transition is much smaller than the universe's radiation energy. However, the supercooling FOPT favored in our scenario is the delayed onset of the nucleation compared to the critical point, i.e. $T_n \ll T_c$, which consequently produces large $\Delta U(T_n)$. The released energy can be as large as the universe's energy but should not reheat the universe too much; otherwise, Fermi-balls can be melted and destroyed.

The size of the bubbles are negligible right after nucleation, but they expand quickly with the wall velocity $v_b$. As a result, the volume occupied by the old phase decreases with temperature. This can be quantitatively described by the fraction of the volume that remains in the old phase~\cite{Guth:1981uk}
\be\label{pT}
p(T)=e^{-I(T)},
\ee
where the exponent $I(T)$ is defined as 
\be\label{IT}
I(T)=\frac{4\pi}{3}\int_{T}^{T_c}dT'\frac{\Gamma(T')}{T'^4H(T')}\left[\int_{T}^{T'}d\tilde T\frac{v_b}{H(\tilde T)}\right]^3.
\ee
It is expected that $p(T)\to 0$ as the FOPT proceeds and completes.

The temperature that the bubbles are sufficient to form an infinite connected cluster is called the percolation temperature $T_p$, which satisfies $p(T_p)=0.71$~\cite{rintoul1997precise}, corresponding to $I(T_p)=0.34$. This is also commonly adopted as the temperature at which FOPT GWs are produced~\cite{Megevand:2016lpr,Kobakhidze:2017mru,Ellis:2018mja,Ellis:2020awk,Wang:2020jrd}. 

After percolation, the fraction of old phase remnants keeps decreasing. Based on the numerical result in Ref.~\cite{rintoul1997precise}, we define the lowest temperature at which ``old phase remnants can still form an infinite connected cluster'' as $T_*$, which satisfies $p(T_*)=1-p(T_p)=0.29$, corresponding to $I(T_*)=1.24$. 

$T_*$ is also the temperature at which Fermi-balls start to form. Just below $T_*$, the old phase still occupies a sizable fraction $\sim 0.29$ of the whole universe, but it is separated in many ``false vacuum remnants''. Those remnants might split into smaller pieces before they eventually shrink to tiny size to be Fermi-balls. The critical size $R_*$ of a remnant at the end of the splitting (and hence at the beginning of the shrinking) is the one that shrinks to negligible size before another bubble containing the true vacuum is created inside it~\cite{Krylov:2013qe}. This means
\be\label{right_after}
\Gamma(T_*)V_*\Delta t\sim1,\quad V_*=\frac{4\pi}{3}R_*^3,
\ee
where $\Delta t=R_*/v_b$ is the time cost for shrinking. The number density of the remnants $n_{\rm FB}^*$ satisfy $n_{\rm FB}^*V_*=p(T_*)$ so that it can be written as
\be
n_{\rm FB}^*=\left(\frac{3}{4\pi}\right)^{1/4}\left(\frac{\Gamma(T_*)}{v_b}\right)^{3/4}p(T_*).
\ee
The $Q$-charge trapped in a remnant is
\be
Q_{\rm FB}^*=F_\chi^{\rm trap.}\frac{c_\chi\eta_Bs_*}{n_{\rm FB}^*}.
\ee
Since $n_{\rm FB}/s$ and $Q_{\rm FB}$ do not change during the adiabatic evolution of the universe, at present universe they are
\be\label{nQ_profile}
n_{\rm FB}=\frac{n_{\rm FB}^*}{s_*}s_0, \quad Q_{\rm FB}=Q_{\rm FB}^*,
\ee
where $s_0=2891.2~{\rm cm}^{-3}$ is the cosmic entropy density today~\cite{Tanabashi:2018oca}.

%%%%%
\subsection{Stability and profile of Fermi-ball} \label{sec:profile}

At present universe ($T\approx0$)~\footnote{After pair annihilation, the remaining $\chi$ fermions are still very hot but out of equilibrium. They can cool down by emitting light SM particles such as electrons/neutrons/photons (through the $\phi$-portal couplings) via the Fermi-ball surface. Following Ref.~\cite{Witten:1984rs}, we confirm that the cooling time scale is $\lesssim10^{-3}\times$ the Hubble time, thus such cooling is efficient and Fermi-ball can reach $T\sim0$ today.}, the energy of a Fermi-ball with global charge $Q_{\rm FB}$ and radius $R$ is~\cite{Lee:1986tr}
\be\label{FB_E}
E=\frac{3\pi}{4}\left(\frac{3}{2\pi}\right)^{2/3}\frac{Q_{\rm FB}^{4/3}}{R}+4\pi\sigma_0 R^2+\frac{4\pi}{3}U_0R^3,
\ee
where the first term is the Fermi-gas pressure of the $\chi$ constituents, $\sigma_0$ the surface tension, and $U_0 \equiv \Delta U(T)|_{T=0}$. In our scenario, the surface term is negligible compared to the volume one because a Fermi-ball turns out to be of macroscopic size. By minimizing $E$ with respect to the radius, i.e., by solving $dE/dR|_{R_{\rm FB}}=0$, we obtain the mass and radius of a Fermi-ball
\bea\label{M_FB_1}
M_{\rm FB}=&&~E\big|_{R=R_{\rm FB}}=Q_{\rm FB}\left(12\pi^2U_0\right)^{1/4},\nn\\
R_{\rm FB}=&&~Q_{\rm FB}^{1/3}\left[\frac{3}{16}\left(\frac{3}{2\pi}\right)^{2/3}\frac{1}{U_0}\right]^{1/4}.
\eea
The Fermi-ball is stable if
\be\label{stability}
\frac{dM_{\rm FB}}{dQ_{\rm FB}}<M_\chi\equiv g_\chi w_0,\quad \frac{d^2M_{\rm FB}}{dQ_{\rm FB}^2}\leqslant 0,
\ee
where $w_0\equiv w(T)|_{T=0}$. The first condition implies that a $\chi$ has smaller energy inside the Fermi-ball than outside; and the second one implies that the $\chi$'s energy inside the ball becomes smaller for a larger total charge, energetically favoring a larger ball for a given total charge or being stable against the fission into smaller balls. The second condition is automatically satisfied.

Given a scalar potential $U(\phi,T)$, it is convenient to rewrite Fermi-ball profiles in Eqs.~(\ref{nQ_profile}) and (\ref{M_FB_1}) in terms of the action at $T_*$. Since $\Gamma(T_*)/T_*^4\approx e^{-S_3(T_*)/T_*}$, the mass, radius and charge of a Fermi-ball are rewritten as
\begin{multline}\label{estimate_MFB}
M_{\rm FB}\approx4.84\times10^{11}~{\rm kg}
\times F_\chi^{\rm trap.}\left(\frac{c_\chi}{0.0146}\right)\left(\frac{U_0^{1/4}}{100~{\rm GeV}}\right)\\
\times\left(\frac{v_b}{0.6}\right)^{3/4}\exp\left\{\frac34\left(\frac{S_3(T_*)}{T_*}-140\right)\right\},
\end{multline}
\begin{multline}
R_{\rm FB}\approx1.08\times10^{-6}~{\rm m} \times \left(F_\chi^{\rm trap.}\frac{c_\chi}{0.0146}\right)^{1/3}\left(\frac{100~{\rm GeV}}{U_0^{1/4}}\right)\\
\times\left(\frac{v_b}{0.6}\right)^{1/4}\exp\left\{\frac14\left(\frac{S_3(T_*)}{T_*}-140\right)\right\},
\end{multline}
\begin{multline} \label{estimate_QFB}
Q_{\rm FB}\approx8.26\times10^{35}\times F_\chi^{\rm trap.}\left(\frac{c_\chi}{0.0146}\right)\left(\frac{v_b}{0.6}\right)^{3/4}\\
\times\exp\left\{\frac34\left(\frac{S_3(T_*)}{T_*}-140\right)\right\}.
\end{multline}
The number density of Fermi-balls is
\begin{multline}\label{estimate_nFB}
n_{\rm FB}\approx4.60\times10^{-39}~{\rm m}^{-3}\times\left(\frac{0.6}{v_b}\right)^{3/4}\\
\times\exp\left\{-\frac34\left(\frac{S_3(T_*)}{T_*}-140\right)\right\}.
\end{multline}
The normalization factor ``0.0146'' for $c_\chi$ will be explained very soon in the next subsection. The number ``140'' used in the exponent is motivated by that $S_3(T_*)/T_* = 140$ is the typical nucleation threshold for an electroweak scale FOPT in a radiation-dominated universe~\cite{Quiros:1999jp,Grojean:2006bp}. However, as emphasized, the released energy can be important in our scenario, so that the threshold $S_3(T_*)/T_*$ can be sizably different from 140.  As $\left(S_3(T_*)/T_*-140\right)$ is in the exponent, its deviation from 0 gives a huge impact on Fermi-ball profiles. The pre-factors in Eqs.~(\ref{estimate_MFB})--(\ref{estimate_nFB}) can be poor estimates for Fermi-balls; rather, for a given model, one should derive $S_3(T_*)/T_*$ to get the real values of the profile.

Finally, as an aside, it is useful to compare the Fermi-ball with the well-known $Q$-ball. $Q$-balls also localize conserved charges in small objects but are made of scalar particles. As a result, they have $M_{\rm QB}\propto Q_{\rm QB}^{3/4}$ and $R_{\rm QB}\propto Q_{\rm QB}^{1/4}$~\cite{Krylov:2013qe}, different from \Eq{M_FB_1}. This means that a Fermi-ball is typically heavier and larger than a $Q$-ball for a given amount of localized $Q$-charge. It can be understood as fermions tend to occupy larger space and more excited energy states due to the Pauli exclusion principle; indeed, the difference technically comes from the different quantum pressure term of the ground-state condensation of scalar particles $\propto Q_{\rm QB}/R$ compared to the fermion's $Q_{\rm FB}^{4/3}/R$ in \Eq{FB_E}. In other words, the mass density of a single Fermi-ball
\begin{multline}
M_{\rm FB}\left(\frac{4\pi}{3}R_{\rm FB}^3\right)^{-1} \, = \, 9.15\times10^{28}~{\rm kg/m}^3\\
\times\left(\frac{U_0^{1/4}}{100~{\rm GeV}}\right)^4,
\end{multline}
is typically smaller than that of a $Q$-ball with $\sim10^{36}~{\rm kg/m}^3$~\cite{Krylov:2013qe}. 
Although both are much denser than a neutron star with a density $\sim 10^{17}~{\rm kg/m}^3$, Fermi-balls and $Q$-balls are not black holes as they are much larger than their Schwarzschild radii; for example, the Schwarzschild radius of a Fermi-ball with $M_{\rm FB} = 10^{11}$ kg $\simeq 10^{-19} M_\odot$ in \Eq{estimate_MFB} is only $\sim 10^{-16}$ m $\ll R_{\rm FB} = 10^{-6}$ m.

%%%%%
\subsection{Fermi-ball DM abundance} \label{sec:FBDM}

The relic density of Fermi-balls is
\be
\Omega_{\rm FB}h^2=\frac{n_{\rm FB}M_{\rm FB}}{\rho_{\rm c}}h^2,
\ee
where $\rho_{\rm c}=3H_0^2M_{\rm Pl}^2/(8\pi)$ and $H_0$ are respectively the critical energy density and Hubble constant today, and $h=H_0/(100~{\rm km}\cdot {\rm s}^{-1}\cdot{\rm Mpc}^{-1})$. Using \Eq{nQ_profile} and \Eq{M_FB_1}, we obtain
\be\label{FB_abundance}
\Omega_{\rm FB}h^2=0.12\times F_\chi^{\rm trap.} \left(\frac{c_\chi }{0.0146}\right) \left(\frac{U_0^{1/4}}{100~{\rm GeV}}\right),
\ee
where we have used the observed DM relic density $\Omega_{\rm DM}h^2=0.12$~\cite{Aghanim:2018eyx,Tanabashi:2018oca} to normalize the expression. Notably, the exponential factors cancel out due to the $Q_{\rm FB}$ dependence of $M_{\rm FB}$ in \Eq{M_FB_1}, a particular result of the fermion nature of constituents; scalar $Q$-balls do not show this. After all, to explain the full DM abundance, we must have $F_\chi^{\rm trap.}\sim1$ and $c_\chi U_0^{1/4}\sim1.46$ GeV. Obviously, this is possible in a large range of parameter space, in principle up to the Planck scale.

In addition to \Eq{FB_abundance} from Fermi-balls, however, there are other contributions to the dark matter relic density from free $\chi$ fermions outside Fermi-balls, i.e. in the true vacuum. The free $\chi$ contributions consist of two parts: one is the fermions that escape from the false vacuum with $F_\chi^{\rm trap}<1$ and hence asymmetric in $\chi$ and $\bar\chi$, and the other is the thermally produced fermions via process $\phi\phi\to\chi\bar\chi$ and hence symmetric.
We have checked that the interaction $-g_\chi\bar\chi\chi\phi$ is always in thermal equilibrium for $g_\chi\gtrsim1$ and $M_\chi^*/T_*\lesssim25$. If this thermal contribution is dominant, the free $\chi$'s will experience the normal freeze-out process, yielding a relic abundance~\cite{Bertone:2004pz}
\bea\label{freeze_out}
(\Omega_{\chi}^{\rm free}+\Omega_{\bar\chi}^{\rm free})h^2\approx&&~\frac{2.55\times10^{-10}~{\rm GeV}^{-2}}{\ave{\sigma v}}\nn\\
\approx&&~0.11\times\frac{1}{g_\chi^4}\left(\frac{M_\chi}{1~{\rm TeV}}\right)^2.
\eea
If, however, the escaping part is dominant, the relic abundance is given by the excess of $\chi$ over $\bar\chi$,
\bea \label{escaping}
\Omega_{\chi}^{\rm free}h^2=&&~(1-F_\chi^{\rm trap.})c_\chi\eta_Bs_0M_\chi\\
=&&~0.036
\times\left(\frac{1-F_\chi^{\rm trap.}}{0.1}\right)\left(\frac{c_\chi}{0.0146}\right)\left(\frac{M_\chi}{1~{\rm TeV}}\right),\nn
\eea
and $\Omega_{\bar\chi}^{\rm free}h^2=0$.

%%%%%%%
\section{A toy model: $\phi^3$-induced FOPT} \label{sec:toy}

In this section, we work out and demonstrate a Fermi-ball DM scenario using a toy model. 
In addition to the basic setup in Section~\ref{sec:setup}, the model has a $\phi$ potential
\be\label{TBR}
U(\phi,T)=\frac12(\mu^2+c\,T^2)\phi^2+\frac{\mu_3}{3}\phi^3+\frac{\lambda}{4}\phi^4,
\ee
where $c>0$ denotes the thermal mass correction from light degrees of freedom, e.g. $\phi$ itself (heavier field contributions such as $\chi$ are Boltzmann suppressed), while thermal corrections to other terms are assumed to be smaller than the tree-level ones. If $\mu^2>0$, $\mu_3<0$ and $\lambda>0$, the potential has two local minima at $T=0$
\be
\ave{\phi} =0,\quad \ave{\phi}=w_0=\frac{1}{2\lambda}\left(-\mu_3+\sqrt{\mu_3^2-4\lambda\mu^2}\right),
\ee
separated by a tree-level induced barrier from the renormalizable operator $\phi^3$~\cite{Chung:2012vg}. Using $w_0$ and $M_\phi^2\equiv d^2U(\phi,T)/d\phi^2|_{T=0,\phi=w_0}$, we rewrite $\mu^2$ and $\lambda$ as
\be
\mu^2=-\frac12\left(M_\phi^2+\mu_3 w_0\right),\quad \lambda=\frac{1}{2w_0}\left(\frac{M_\phi^2}{w_0}-\mu_3\right).
\ee
Note that the positive $\mu^2$ grows with the negative $\mu_3$, which is considered in the range
\be \label{mu3_region}
-\frac{3M_\phi^2}{w_0}<\mu_3<-\frac{M_\phi^2}{w_0},
\ee
where the upper limit is for a positive $\mu^2$ so that the barrier still exists at $T=0$ (actually, this is not necessary, but is useful for a large $w_*/T_*$), while the lower limit ensures a global minimum at $\phi=w_0$.

The critical temperature and VEV are
\bea
T_c&=&\frac{1}{3\sqrt{2c}} \sqrt{\frac{9 M_\phi^4-\mu_3^2 w_0^2}{M_\phi^2-\mu_3 w_0}},\nn\\
w_c&\equiv&w(T_c)=T_c\frac{-4\mu_3 w_0^2}{3 M_\phi^2-3 \mu_3 w_0}.
\eea
Although $T_c$ exists for any $\mu_3$ in the range \Eq{mu3_region}, it turns out that nucleation may fail if $|\mu_3|$ is too large.
The action of the bounce solution of the potential in \Eq{TBR} can be derived analytically~\cite{Dine:1992wr}
\be\label{S3T}
\frac{S_3(T)}{T}=\frac{123.48(\mu^2+c\,T^2)^{3/2}}{2^{3/2}T\mu_3^2}f\left(\frac{9(\mu^2+c\,T^2)\lambda}{2\mu_3^2}\right),
\ee
where
\be
f(u)=1+\frac{u}{4}\left(1+\frac{2.4}{1-u}+\frac{0.26}{(1-u)^2}\right).
\ee

\begin{figure}
\centering
\subfigure{
\includegraphics[scale=0.4]{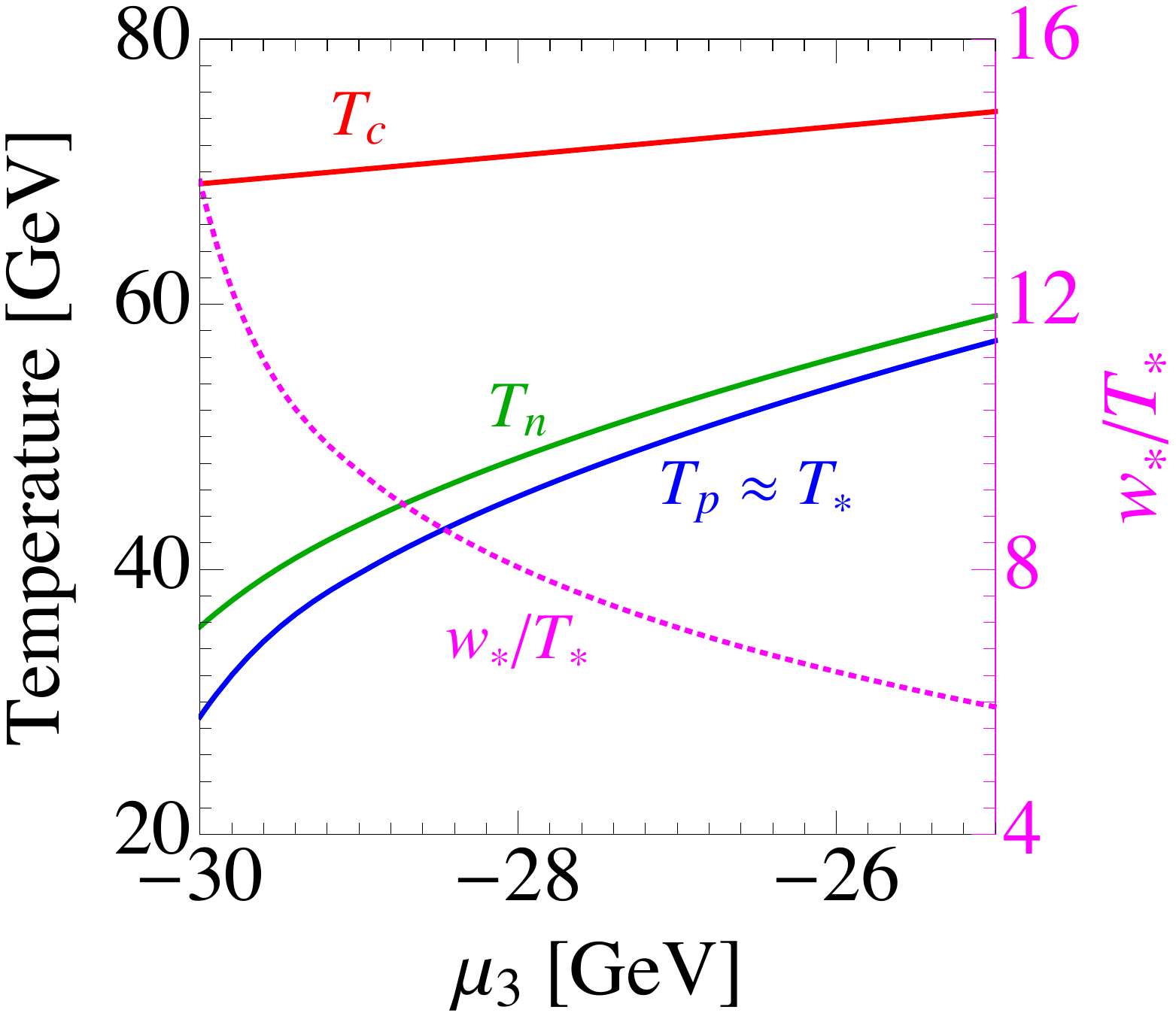}}
\caption{The characteristic temperatures $T_c, T_n, T_p$, and $T_*$ of the FOPT in the benchmark model with parameters in \Eq{benchmark}. The supercooling is apparent from the delayed onset of the nucleation, i.e. $T_n \ll T_c$. But for $\mu_3 \lesssim -30$ GeV, nucleation fails although $T_c$ exists. Also shown in the magenta dashed curve is $w_*/T_*$. $v_b=0.6$ is adopted in the calculation.}
\label{fig:FOPT_numerical}
\end{figure}

For a numerical study, a particularly interesting parameter space that can explain the full DM abundance is (from \Eq{FB_abundance})
\be\label{GWbenchmark}
U_0\sim(100~{\rm GeV})^4, \quad c_\chi\sim0.01.
\ee
As will be discussed in the next section, the 100 GeV scale can typically yield stochastic GWs at milli-Hz frequency, relevant to the future space-based missions. Lower scales may be constrained by Higgs physics and various LHC searches, while higher scales may require too small $c_\chi$ for the DM abundance. Further, we focus on the following benchmark parameters 
\be \label{benchmark}
w_0=400~{\rm GeV},\quad M_\phi=100~{\rm GeV},\quad c=0.4,
\ee
which yield $U_0 \simeq (100 \,{\rm GeV})^4$.
We scan $\mu_3$ over the range in \Eq{mu3_region}, which is numerically $[-75,-25]$ GeV for this benchmark. 
But by plugging \Eq{S3T} into \Eq{nucleation_condition}, we find that nucleation $T_n$ cannot be solved for $\mu_3 \lesssim-30$ GeV, partly because the potential barrier becomes too high. Then using \Eq{pT} and (\ref{IT}), we obtain the fraction of the false vacuum volume $p(T)$, and derive $T_p$ and $T_*$ by requiring $p(T)=0.71$ and 0.29~\footnote{The physical volume of the false vacuum, $a^3(T)p(T)$ with $a(T)$ being the scale factor in the Robertson-Walker metric, eventually decreases to zero with temperature, completing the FOPT~\cite{Turner:1992tz}.}. 

Fig.~\ref{fig:FOPT_numerical} shows the resulting characteristic temperatures $T_c$, $T_n$, $T_p$ and $T_*$ in the benchmark as functions of $\mu_3$. The supercooling is apparent from the large difference between $T_c$ and $T_n$, i.e., the delayed onset of nucleation. Its strength increases with $|\mu_3|$ again because it yields higher barrier and deeper true minimum.

In the same Fig.~\ref{fig:FOPT_numerical}, also plotted in the magenta curve is $w_*/T_*$. It ranges from 6 to 14 for the given range of $\mu_3$ so that $g_\chi\sim\mO(1)$ is sufficient for $M_\chi^*=g_\chi w_*\gg T_*$ (hence, large $F_\chi^{\rm trap.}$); $w_*/T_*$ also measuring the strength of the FOPT grows with $|\mu_3|$, which is consistent with more delayed onset of nucleation.
The first stability condition in \Eq{stability}, $g_\chi w_0>(12\pi^2 U_0)^{1/4}$, is also satisfied for $g_\chi \gtrsim 0.82$.  Thus, the supercooling with reasonable $g_\chi \sim 1$ can produce stable Fermi-balls. Their properties in Eqs.~(\ref{estimate_MFB})--(\ref{estimate_nFB}) vary in the range (for $\mu_3 = -30 \sim -25$ GeV)
\bea
n_{\rm FB}&=&1.1\times10^{-37}~{\rm m}^{-3}\sim 9.3\times10^{-34}~{\rm m}^{-3},\nn\\
Q_{\rm FB}&=&3.9\times10^{34}\sim 4.0\times10^{30},\nn\\
M_{\rm FB}&=&2.4\times10^{10}~{\rm kg}\sim2.6\times10^6~{\rm kg},\nn \\
R_{\rm FB}&=&3.7\times10^{-7}~{\rm m}\sim1.8\times10^{-8}~{\rm m},
\label{FBproperties} \eea
which are only a few orders of magnitudes different from the pre-factors in Eqs.~(\ref{estimate_MFB})--(\ref{estimate_nFB}). They are macroscopic compared to individual constituents. Their astrophysical signals will be discussed in the next section.

\begin{figure}
\centering
\subfigure{
\includegraphics[scale=0.37]{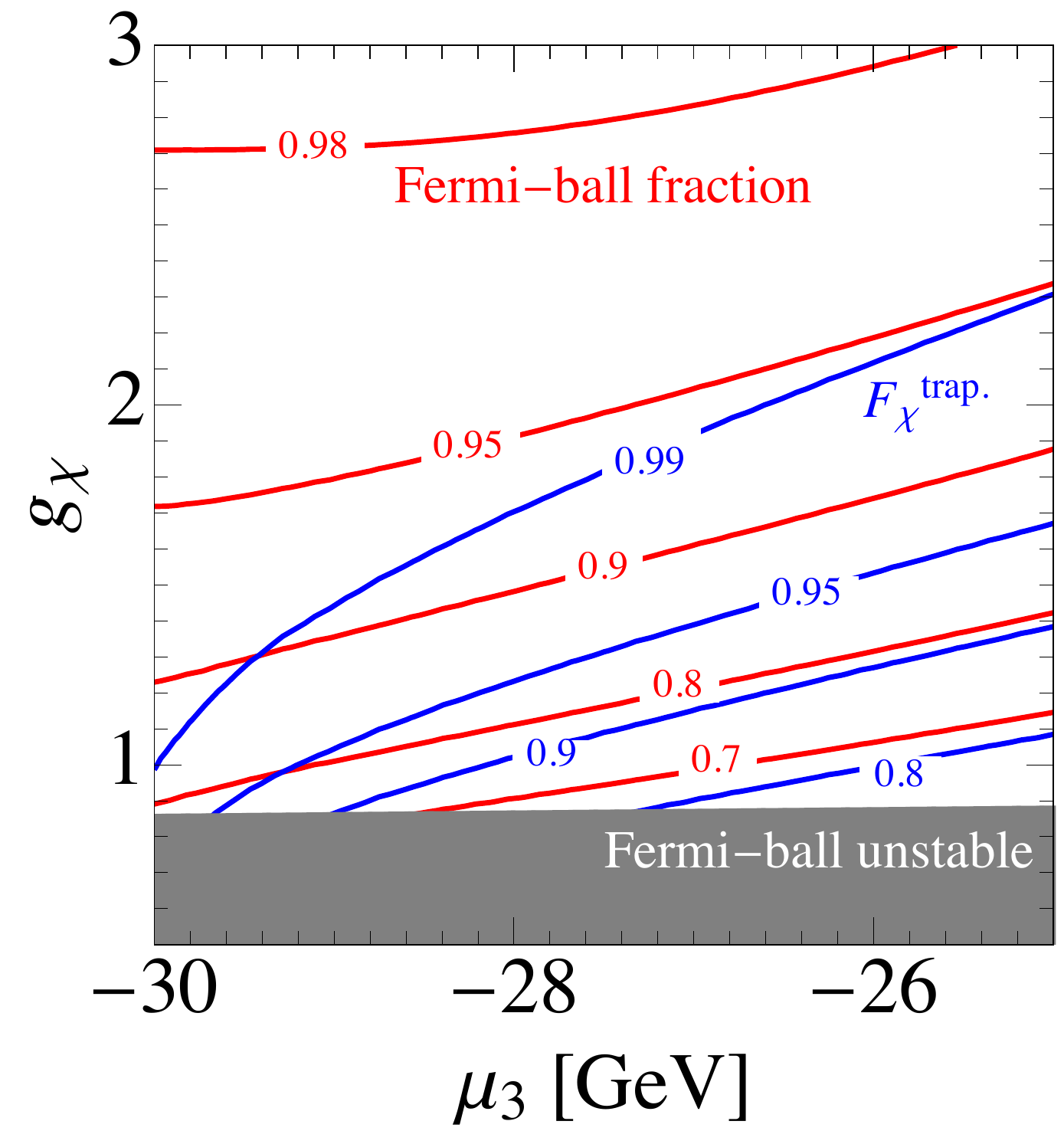}}
\caption{The fraction of Fermi-ball's contribution to the total DM abundance (red solid). The total abundance $\Omega_{\rm tot} h^2$ from Fermi-balls and free $\chi$'s is fixed to 0.12 by choosing a proper $c_\chi$. Also overlaid are $F_{\chi}^{\rm trap.}$ contours (blue solid) for reference. For $g_\chi \lesssim 0.82$, Fermi-balls are unstable. $v_b=0.6$ is adopted in the calculation.}
\label{fig:FBfrac}
\end{figure}

The benchmark Fermi-balls can explain the full abundance $\Omega_{\rm FB}h^2=0.12$ for some $c_\chi$ near $0.01$. Accounting for subdominant contributions from free $\chi$'s and choosing a proper $c_\chi$ to set the total abundance $\Omega_{\rm tot}h^2 = \Omega_{\rm FB}h^2 + \Omega_{\chi}^{\rm free}h^2 + \Omega_{\bar \chi}^{\rm free}h^2 =0.12$, we show the fraction of Fermi-ball's contribution $\Omega_{\rm FB} / \Omega_{\rm tot}$ in Fig.~\ref{fig:FBfrac} as a function of $\mu_3$ and $g_\chi$. For $g_\chi \gtrsim 0.82$ with stable Fermi-balls, the Fermi-ball fraction is generally high above $80\sim90\%$. It increases with $F_\chi^{\rm trap.}$ (equivalently with $|\mu_3|$) because the escaping contribution in \Eq{escaping} becomes smaller; and increases also with $g_\chi$ as the thermal contribution in \Eq{freeze_out} is suppressed by efficient annihilations.

%%%%%%%
\section{Signals of Fermi-ball DM} \label{sec:GW}

%%%
\subsection{Absence of DM signals}

The number density of Fermi-balls is extremely small. 
With $n_{\rm FB}=10^{-39}~{\rm m}^{-3}$ from \Eq{estimate_nFB}, the number of Fermi-balls passing through a detector with size $L=10$ m is only $n_{\rm FB}v_{\rm DM}L^2 \sim10^{-24}$/year, where $v_{\rm DM}\sim10^{-3}$ is the virial velocity of a galaxy.  Even considering a reasonable exponential factor in \Eq{estimate_nFB}, it is unlikely to observe Fermi-balls in any direct detection experiments. From another point of view, this can explain the null results from direct detection experiments so far~\footnote{Although there have been several reports of possible detections such as the most recent XENON1T excess of the electron recoil in the keV region~\cite{Aprile:2020tmw}, their verifications as well as DM origins are under serious disputes both experimentally and theoretically.}.
On the other hand, the free $\chi$ may have direct detection signals if $\phi$ has a portal mixing with the SM Higgs boson. The spin-independent $\chi$-nucleon cross section is
\be
\sigma^{\rm SI}_{\chi p}\approx6\times10^{-45}~{\rm cm}^2\times\left(\frac{1-F_\chi^{\rm trap.}}{0.01}\right)g_\chi^2\sin^2\alpha,
\ee
where $\alpha$ is the mixing angle between $\phi$ and $h$. We have used the formulae in~\cite{Li:2014wia} and the nucleon form factors from~\cite{Belanger:2013oya}. The collider experiments have constrained $\sin^2\alpha\lesssim10^{-1}$~\cite{Aad:2019mbh}. If $g_\chi$ is sizable, $\sigma^{\rm SI}_{\chi p}$ might reach the direct detection limit, which is $\sim10^{-45}~{\rm cm}^2$ for $M_\chi\gtrsim1$ TeV~\cite{Schumann:2019eaa}. However, increasing $g_\chi$ will decrease $(1-F_\chi^{\rm trap.})$ and hence suppress $\sigma^{\rm SI}_{\chi p}$, as shown in Fig.~\ref{fig:FBfrac}.
  
Furthermore, although macroscopic in size and mass compared to constituent particles, Fermi-balls cannot induce interesting astrophysical signals either. They are still much lighter ($10^{-20} M_\odot \sim 10^{-24} M_\odot$ in \Eq{FBproperties}) than even most of the largest asteroids in the solar system, while being much farther away. They are not so compact (much bigger than their Schwarzschild radii) that their gravitational effects are diluted. They are so sparsely distributed that their accidental coalescence or confront with other astrophysical structures are rare. Therefore, no astrophysical signals are generally expected.
  
It is unlikely that Fermi-ball itself produces detectable signals. In this section, we investigate a detectable signal from a FOPT -- the stochastic phase transition GWs -- in the toy model.

%%%%%
\subsection{Stochastic GW} \label{sec:FBGW}

\medskip
A FOPT produces stochastic GWs via bubble collisions, sound waves and turbulence in the plasma~\cite{Mazumdar:2018dfl}. GWs are assumed to be produced at the percolation temperature $T_p$. In general, the GW energy density spectrum $\Omega_{\rm GW}(f)=\rho_c^{-1}d\rho_{\rm GW}/d\ln f$ (with $\rho_{\rm c,GW}$ being the universe's critical and GW energy densities) can be expressed as numerical functions of two effective parameters~\cite{Grojean:2006bp,Caprini:2015zlo,Caprini:2019egz}: (i) the ratio of the phase transition latent heat to the universe's radiation energy density
\be
\alpha=\left(\Delta U(T_p)+T\frac{\partial U(\phi,T)}{\partial T}\Big|_{T_p}\right) \, \rho_R^{-1}(T_p),
\ee
(ii) the inverse ratio of the time scales of the FOPT and the Hubble expansion, $\beta/H(T_p)$, where
\be
\beta \, = \,  H(T_p) T_p \frac{d}{dT}\left(\frac{S_3(T)}{T}\right)\Big|_{T=T_p}.
\label{beta} \ee
$\alpha$ is related to the strength of the GWs, while $\beta$ to the inverse duration of the FOPT,  hence the characteristic frequency of the GWs.

This general description, however, requires a modification in our case with significant supercooling.
First, since the $\phi$'s vacuum energy is important in the Hubble constant during a FOPT, the radiation dominance and adiabatic expansion assumed above may not be exact. In addition, since $S_3(T)/T$ may change rapidly with $T$, expanding this linearly in $T$ around $T_p$ may not be a good approximation. It was suggested that $\beta/H(T_p)$ must be replaced more generally by~\cite{Megevand:2016lpr,Kobakhidze:2017mru,Ellis:2018mja,Ellis:2020awk,Wang:2020jrd}
\be
\frac{\beta}{H(T_p)}\to\frac{(8\pi)^{1/3}}{H(T_p)\bar R}v_b,
\ee
where $\bar R$ is some relevant length scale of bubbles~\cite{Hindmarsh:2017gnf}, which in this study we use the mean separation of bubbles $R_p$ at $T_p$~\cite{Wang:2020jrd}. In all, we use formulae in \cite{Grojean:2006bp,Caprini:2015zlo} with this replacement to calculate $\Omega_{\rm GW}(f)$, and take the finite duration of sound waves period into account. The energy budget of FOPT is calculated using the numerical results in Ref.~\cite{Espinosa:2010hh}.

It turns out that GW peaks are determined by sound waves, while high-frequency tails are modified by turbulence. Since the bubble wall reaches its terminal velocity $v_b$ rapidly, only a tiny fraction of released energy is transferred to the wall, making the bubble collision contribution negligible. Rather, most energy pumped into the fluid around bubbles makes the sound wave a dominant source of the GWs~\cite{Ellis:2018mja}. As the sound wave period usually lasts shorter than a Hubble time~\cite{Ellis:2018mja,Ellis:2020awk,Wang:2020jrd,Schmitz:2020rag,Guo:2020grp}, after that non-linear fluid motions can source further GWs via turbulence.

\begin{figure}
\centering
\subfigure{
\includegraphics[scale=0.35]{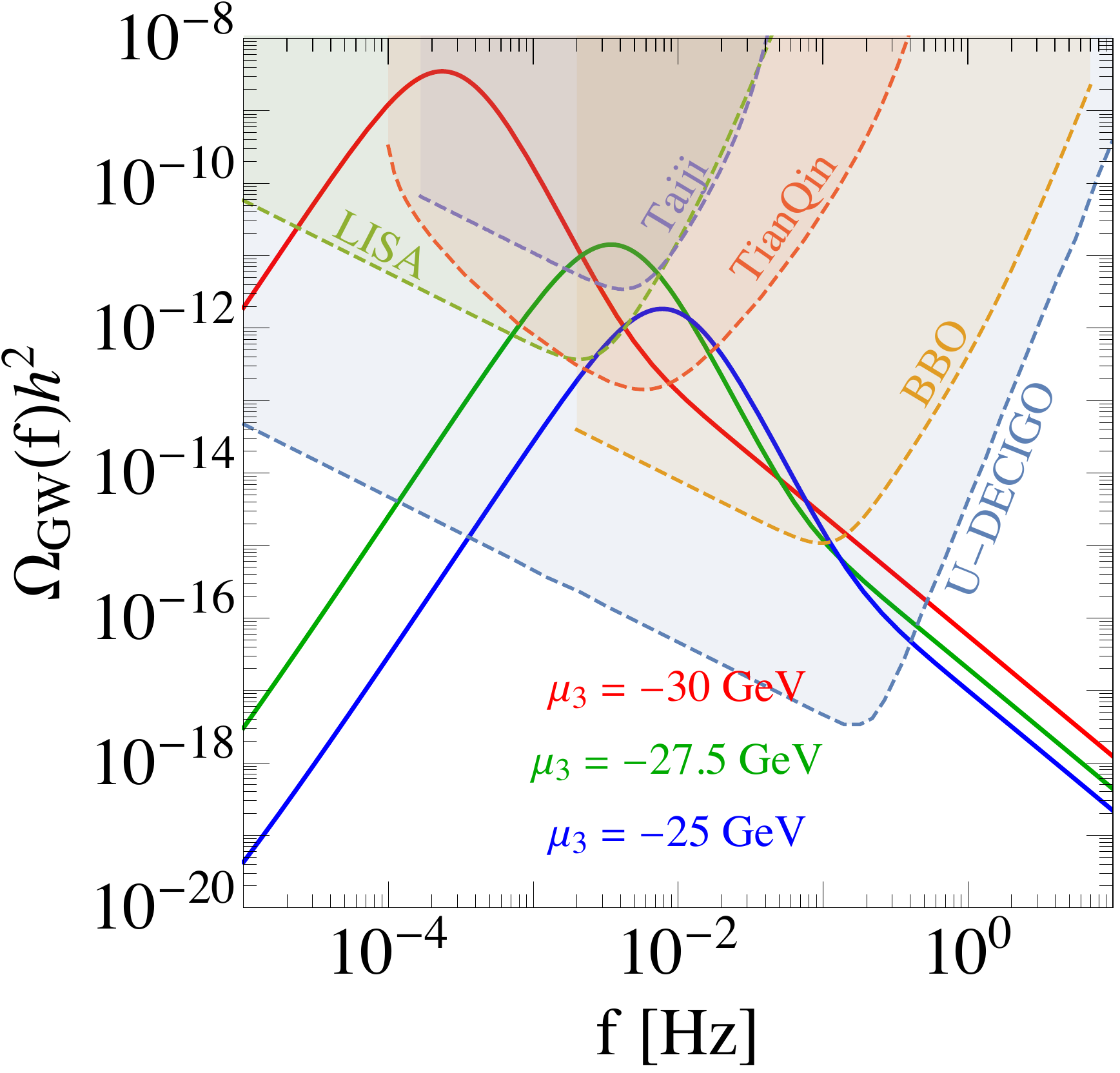}}
\caption{FOPT GW spectra $\Omega_{\rm GW}(f)$ in the benchmark model with \Eq{benchmark} for three chosen values of $\mu_3 = -30, -27.5$ and $-25$ GeV. The sensitivity curves of future detectors are overlaid. $v_b=0.6$. GW peaks are determined by sound waves, while high-frequency tails are modified by turbulence. }
\label{fig:GW}
\end{figure}

\medskip
Fig.~\ref{fig:GW} shows the GW spectra for the benchmark parameters with three chosen values of $\mu_3= -30, -27.5,$ and $-25$ GeV, overlaid with the sensitivity curves of future detectors. Above all, it is clear that observable GW signals at milli-Hz frequencies are possible, for the weak-scale phase transition. The mHz scale is relevant to the next-generation space-based GW detectors such as LISA~\cite{Audley:2017drz}, BBO~\cite{Crowder:2005nr}, TianQin~\cite{Luo:2015ght,Hu:2017yoc}, Taiji~\cite{Hu:2017mde,Guo:2018npi} or DECIGO~\cite{Kawamura:2011zz,Kawamura:2006up}. GWs are enhanced for larger $|\mu_3|$, as the supercooling is more significant. Quantitatively, for $\mu_3=-30 \sim -25$ GeV, the parameter $\alpha=6.2\sim0.6$ (varying from ultra supercooling to strong supercooling as classified in Ref.~\cite{Wang:2020jrd}). These are not negligible in the Hubble expansion rate near the phase transition temperature as discussed, but they reheat the universe only by a factor of $(1+\alpha)^{1/4} \lesssim 1.6$, hence safe. For larger $\mu_3>-25$ GeV, although the FOPT weakens with smaller $w_*/T_*<6$ producing weaker GWs, the Fermi-ball DM scenario can still work with a stronger (but perturbative) $g_\chi$.

Also can be seen in the figure is that the stronger the supercooling is, the later the GW is produced, hence at the lower frequencies. The peak frequency can be estimated in the case of FOPTs with $\alpha \lesssim 1$. From \Eq{nucleation_condition}, the nucleation/percolation happens when $S_3(T_p)/T_p \simeq 4 \ln g_p^{1/2} M_{\rm Pl}/T_p \sim 100$ for $T_p \sim 100$ GeV with radiation-dominance. If $S_3(T)/T$ varies slowly (or, logarithmically) with $T$ near $T_p$, \Eq{beta} yields $\beta/H(T_p) \sim S_3(T_p)/T_p \sim 100$.
With the characteristic GW frequency $f_p \sim R_p \sim \beta / v_b$, this can be translated into the peak frequency observed today~\cite{Grojean:2006bp,Caprini:2015zlo},
\bea\label{frequency}
f_0 \,&=&\, f_p \frac{a_p}{a_0} \,\sim \, \frac{\beta}{v_b} \left( \frac{g_{0}}{g_{p}}\right)^{1/3} \frac{T_0}{T_p} \nonumber\\
 &\simeq&\, \frac{1}{v_b}\left(\frac{g_p}{100}\right)^{1/6} \frac{T_p}{100 \,{\rm GeV}} \,\frac{\beta/H(T_p)}{100} \,{\rm mHz}.
\eea
Fig.~\ref{fig:GW} shows that such estimation is still approximately good for relatively strong supercooling cases with $\mu_3 = -27.5$ and $-25$ GeV. But for the ultra supercooling with $\mu_3 = -30$ GeV, the estimation breaks down more severely (radiation-dominance, adiabatic expansion, and slow variation of $S_3(T)/T$ all may break down) so that the actual $\beta/H (T_p)$ turns out be much smaller ($T_p$ is also smaller), yielding an order-of-magnitudes smaller peak frequency. If the supercooling becomes even stronger with larger $|\mu_3|$, the $\beta/H(T_p) \lesssim 1$ becomes too small to induce nucleation.

The detection of such stochastic GWs, albeit very exciting, does not necessarily imply a Fermi-ball DM scenario. The GW properties depend only on the $\phi$ potential, while Fermi-ball DM scenarios depend additionally on $g_\chi$. Moreover, Fermi-ball DM scenarios can be realized in a much larger parameter space (that may not produce detectable GWs) than considered in this section.

%%%%%%%
\section{Summary} \label{sec:conclusion}

We have developed a new DM scenario, where Fermi-balls formed during a strong FOPT can be the DM candidate. The DM abundance can be explained in a large range of parameter space, determined most crucially by the $\chi$-asymmetry and the FOPT scale through \Eq{FB_abundance}. The necessary conditions and ingredients for Fermi-ball DM have been discussed in the general context and demonstrated in a toy model, so that the mechanism is expected to be applied to a wide varieties of new physics models. 

The Fermi-ball formation has to start with efficient trapping of $\chi$ in the false vacuum, which favors a FOPT with supercooling and a moderate size of $g_\chi$.
Then the Fermi-ball's overall number density is determined solely by the phase transition, while each Fermi-ball's stability and mass arise from the interplay of $\chi$'s quantum pressure and $\phi$'s free energy. Its extensive profiles then exhibit characteristic dependences on $Q_{\rm FB}$ encoding the fermion nature of its constituents, which are thus different from those of $Q$-balls made of scalars. 

The Fermi-ball DM scenario generally produces no detectable signals. Although macroscopic in size and mass, the Fermi-ball is still too small, diffuse and sparsely distributed to induce interesting signals in both terrestrial and astrophysical labs. But in the parameter space with a $\sim100$ GeV energy scale in the FOPT, resulting stochastic GWs will be detectable at the next-generation space-based missions. The detection of GWs, however, does not necessarily imply a Fermi-ball DM scenario. Nevertheless, the discovery of a FOPT by the observation of stochastic GWs would make such a DM scenario more worth considering.

%%%%%%%
\section*{Acknowledgement}

We thank Fa Peng Huang, Ryusuke Jinno, Andrew J. Long and Kengo Shimada for discussions on supercooling, and Lian-Tao Wang for discussions on non-topological DM candidates, and Chang Sub Shin and Dongjin Chway for discussions on FOPT dynamics, and Ran Ding and Bin Zhu for communications on DM direct detections. We are especially grateful to Andrew J. Long for the comments about Fermi-ball cooling. SJ and KPX are supported by Grant Korea NRF-2019R1C1C1010050, and SJ also by POSCO Science Fellowship. JPH is supported by Korea NRF-2015R1A4A1042542 and IBS under the project code, IBS-R018-D1.

%%%%%%%
\appendix
\section{Generating an excess for $\chi$} \label{sec:excess}

The $\chi$-asymmetry can be obtained by using one of the mechanisms reviewed in Refs.~\cite{Kaplan:2009ag,Petraki:2013wwa,Zurek:2013wia}. For example, it can be generated non-perturbatively by the $U(1)_Q$-breaking sphaleron process during the FOPT. After the phase transition, the sphaleron is frozen thus $U(1)_Q$ is conserved again but net $Q$-charge has been accumulated in the universe~\cite{Kaplan:1991ah,Dutta:2006pt,Dutta:2010va,Shelton:2010ta,Petraki:2011mv,Walker:2012ka,Baldes:2017rcu,Gu:2017rzz,Hall:2019rld}. This scenario is analogous to electroweak baryogenesis. On the other hand, there are mechanisms analogous to leptogenesis, where the $\chi$-excess comes from out-of-equilibrium decays of heavy particles; see Refs.~\cite{Cosme:2005sb,An:2009vq,Falkowski:2011xh} and more references in the review~\cite{Zurek:2013wia}.

In this article, we adopt the leptogenesis-style scenario as the benchmark mechanism to get excess of $\chi$. The relevant Lagrangian reads
\bea\label{leptogenesis}
\mL\supset~&&\bar\nu_R^ii\slashed{\partial}\nu_R^i-\sum_j\frac12M_j\left(\overline{\nu_R^c}^j\nu_R^j+{\rm h.c.}\right)\nn\\
&&-\sum_{i,j}\lambda_\nu^{ij}\bar\ell_L^i\tilde H\nu_R^j-\sum_j\lambda_\chi^{j}\bar\chi_LS\nu_R^j+{\rm h.c.},
\eea
where $\nu_R^j$ is the right-handed neutrino, $\ell_L^i$ is the SM left-handed lepton doublet, and $\tilde H$ is the charge conjugate of the SM Higgs. The first term of the second line in \Eq{leptogenesis} is nothing but the standard Yukawa interaction of the leptogenesis mechanism~\cite{Covi:1996wh,Luty:1992un,Asaka:1999yd,Buchmuller:2005eh}, and it gives the decay width asymmetry~\cite{Covi:1996wh}
\bea
\epsilon_\ell=&&~\frac{1}{\Gamma_{\nu_R^1}}\left[\Gamma(\nu_R^1\to\ell H)-\Gamma(\nu_R^1\to\bar\ell H^*)\right]\nn\\
=&&~\frac{1}{\Gamma_{\nu_R^1}}\sum_{k\neq1}
\frac{-3M_{1}}{128\pi^2}\frac{M_{1}}{M_{k}}{\rm Im}\left[(\lambda_\nu^\dagger\lambda_\nu)_{1k}^2\right],
\eea
where we only consider the lightest right-handed neutrino (denoted as $\nu_R^1$), because it dominates the leptogenesis. The width asymmetry comes from the imaginary part of the couplings $\lambda_\nu^{ij}$, which characterizes the $CP$ violating effect.

\begin{figure}
\centering
\subfigure{
\includegraphics[scale=0.8]{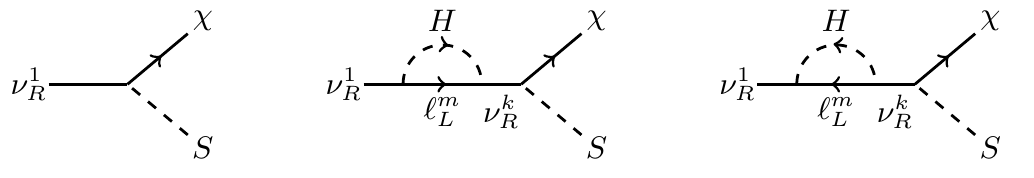}}
\caption{Feynman diagrams generating the width asymmetry of $\nu_R^1\to\chi S$, hence the number asymmetry of $\chi$. $CP$-violating effects come from complex Higgs couplings.}
\label{fig:vR_decay}
\end{figure}

The second term in the second line of \Eq{leptogenesis} involves the interaction among $\nu_R^j$, $\chi_L$ and a new real scalar $S$. It breaks the $U(1)_Q$ explicitly, but thanks to the large mass of the $\nu_R^j$, i.e. $M_j\sim|\lambda_\nu|^2\times3.0\times10^{14}~{\rm GeV}$ required by the seesaw mechanism to provide a left-handed neutrino mass of $\sim0.1$ eV, the breaking of $U(1)_Q$ only happens at this high scale and at low energy it is approximately conserved. For simplicity, we assume the couplings $\lambda_\chi^j$ are real, and all $CP$ violating effects come from $\lambda_\nu^{ij}$. The diagrams relevant to $\nu_R^1\to\chi S$ are shown in Fig.~\ref{fig:vR_decay}, and standard calculation gives the width asymmetry
\bea
\epsilon_\chi=&&~\frac{1}{\Gamma_{\nu_R^1}}\left[\Gamma(\nu_R^1\to\chi S)-\Gamma(\nu_R^1\to\bar\chi S)\right]\\
=&&~\frac{1}{\Gamma_{\nu_R^1}}\sum_{k\neq1}
\frac{-M_{1}}{128\pi^2}\frac{M_{1}}{M_{k}}\lambda_\chi^{1}\lambda_\chi^{k}{\rm Im}\left[(\lambda_\nu^\dagger\lambda_\nu)_{1k}\right]\left(1-\frac{M_S^2}{M_{1}^2}\right)^2,\nn
\eea
where $M_S$ is the mass of $S$. Assuming $|\lambda_\chi^j|\sim|\lambda_\nu^{ij}|$, we get
\be
\epsilon_\chi\approx\frac16\left(1-\frac{M_S^2}{M_{1}^2}\right)^2\epsilon_\ell\equiv c_\chi\epsilon_\ell.
\ee
Consequently,
\be
\eta_\chi=c_\chi\eta_B.
\ee
Thus, the $\chi$-asymmetry is proportional to baryon asymmetry, and the coefficient depends on the mass of $S$. If $M_S$ is close to $M_1$, $c_\chi$ can be fairly small. For instance if $M_S=4M_{1}/5$ then $c_\chi=0.02$. $c_\chi$ can be easily of $\sim$ 0.01 relevant to this paper, depending on $M_S$ as well as the size of various $\lambda_{\nu, \chi}$.

Another point is that the vertex $\bar\chi_L\phi\nu_R^j$ must be forbidden, otherwise $\chi$ can decay to the SM particles via an off-shell $\nu_R^j$ and the $\phi$-portal interactions, and hence the Fermi-ball will disappear. A solution is to assign a $\Z_2$ symmetry under which $\chi$ and $S$ are oddly charged while $\phi$, $\nu_R$ and all SM particles are evenly charged. As long as $M_S>M_\chi$, $\chi$ is the end of the decay chain of the $\Z_2$-odd particles, and hence stable. In a supersymmetric model, $S$ might be identified as the superpartner of $\nu_R$ (and $S$ should be complex in this case) and $\Z_2$ as the $R$-parity.

\bibliographystyle{apsrev}
\bibliography{reference}

\end{document}